# Hierarchical domain structures in buckled ferroelectric free sheets

David Pesquera [a,*], Kumara Cordero-Edwards [a], Marti Checa [b], Ilia Ivanov [b], Blai Casals [a,c], Marcos Rosado [a], José Manuel Caicedo [a], Laura Casado-Zueras [d], Javier Pablo-Navarro [d,e], César Magén [e,f], José Santiso [a], Neus Domingo [b], Gustau Catalan [a,g] and Felip Sandiumenge [h,*]

[a]*Catalan Institute of Nanoscience and Nanotechnology (ICN2), CSIC and BIST, Campus UAB, Bellaterra, Barcelona, Catalonia, 08193 Spain*

[b]*Center for Nanophase Materials Sciences, Oak Ridge National Laboratory, Oak Ridge, Tennessee 37831, USA*

[c]*Dept. of Condensed Matter Physics, University of Barcelona, 08028 Barcelona, Spain*

[d]*Advanced Microscopy Laboratory, University of Zaragoza, Zaragoza, Spain*

[e]*Instituto de Nanociencia y Materiales de Aragón (INMA), Universidad de Zaragoza-CSIC, 50018 Zaragoza, Spain*

[f]*Departamento de Física de la Materia Condensada, Universidad de Zaragoza, 50009 Zaragoza, Spain*

[g]*ICREA-Catalan Institution for Research and Advanced Studies, Passeig Lluïs Companys, Barcelona, Catalonia, Spain*

[h]*Institut de Ciéncia de Materials de Barcelona-CSIC, Campus de la UAB, 08193 Bellaterra, Catalonia, Spain*

*Corresponding authors:
E-mail addresses: david.pesquera@icn2.cat (D.Pesquera), felip@icmab.cat (F.Sandiumenge)

Flat elastic sheets tend to display wrinkles and folds. From pieces of clothing down to two-dimensional crystals, these corrugations appear in response to strain generated by sheet compression or stretching, thermal or mechanical mismatch with other elastic layers, or surface tension. Extensively studied in metals, polymers and, — more recently — in van der Waals exfoliated layers, with the advent of thin single crystal freestanding films of complex oxides, researchers are now paying attention to novel microstructural effects induced by bending ferroelectric-ferroelastics, where polarization is strongly coupled to lattice deformation. Here we show that wrinkle undulations in $BaTiO_3$ sheets bonded to a viscoelastic substrate transform into a buckle delamination geometry when transferred onto a rigid substrate. Using spatially resolved techniques at different scales (Raman, scanning probe and electron microscopy), we show how these delaminations in the free $BaTiO_3$ sheets display a self-organization of ferroelastic domains along the buckle profile that strongly differs from the more studied sinusoidal wrinkle geometry. Moreover, we disclose the hierarchical distribution of a secondary set of domains induced by the misalignment of these folding structures from the preferred in-plane crystallographic orientations. Our results disclose the relevance of the morphology and orientation of buckling instabilities in ferroelectric free sheets, for the stabilization of different domain structures, pointing to new routes for domain engineering of ferroelectrics in flexible oxide sheets.

**I. INTRODUCTION**

The macroscopic functionalities in ferroic materials are strongly linked to their domain microstructures. In ferroelectrics, electrostatic and elastic energies are minimized through the formation of ferroelastic domain structures, and the domain dynamics under an applied electric field play a crucial role in processes that are relevant for the technological application of ferroelectrics, such as electromechanical response [1] or electrothermal energy conversion [2, 3]. In analogy to lead-containing complex solid solutions showing morphotropic phase boundaries [4–6], it is found in general that large susceptibilities can be obtained by promoting competing domain and crystallographic structures in chemically simpler ferroelectrics such as $BaTiO_3$ [7–9], opening alternatives for employing lead-free ferroelectrics in applications.

Recent breakthroughs in the fabrication of single-crystal freestanding sheets of complex oxides [10, 11] open the way to use ferroelectric oxides as elements in flexible electronic devices, capable of sustaining bending deformations [12–14]. Further miniaturization of these bendable devices can be achieved by synthesizing 3D nanoarchitectures with curved geometries, where local strain and strain gradients can induce significant changes in materials via electromechanical coupling [15]. In ferroelectric free sheets, curvilinear geometries couple to the electrical polarization, and the strain distribution across the film curvature has been observed to induce continuous electric dipole rotations from in-plane at the tensile strained side to out-of-plane at the compressive strained side [16]. These observations are supported by phase field simulations [17], and represent the lattice accommodation to the bending deformation of the films.

One way to produce curvilinear geometries in the ferroelectric sheets is to attach them to a viscoelastic substrate. The elastic mismatch between the substrate and the film produces wrinkle patterns in which the film adopts a sinusoidal-shape corrugation with alternating concave and convex regions of constant curvature [18], and the spatial distribution of these patterns can be designed to produce exotic domain structures [19]. These works exemplify the possibility to use 3D architectures for domain engineering in oxide ferroelectrics. Elastic thin films can however display a different stress-driven instability when attached to a rigid substrate [20]: if the film is not firmly attached to the substrate, buckle delamination can occur. Contrary to the wrinkles on films attached to viscoelastic substrates where the curvature is constant and the strain is uniformly distributed, the buckle delaminations localize the strain, and curvature is not homogeneous across the profile of the delamination. This buckling deformation therefore represents an alternative curvilinear geometry with potential for domain engineering in ferroelectric sheets, that remains unexplored.

Here we fabricate $BaTiO_3$ free sheets on rigid silicon, showing wrinkles with a geometry closer to that of a buckle delamination. By using several different microscopy techniques to visualize the ferroelectric domains (Optical, Raman, Scanning Probe and Electron microscopy), we identify a complex microstructure contained at these wrinkles, with coexistence of 180º domain patterns at the crest, *c*-oriented domains at the flat slopes and polydomain bundles formed by $a_1/c$ and $a_2/c$ structures, localized at the folded regions of the buckle delamination. Notably, we find that the deviation of the wrinkle with respect to the main crystallographic directions promotes the appearance of additional ferroelastic hierarchical domain structures and increases the domain density of the polydomain bundles. Overall, we observe that the change of geometry from sinusoidal wrinkles where bending strain is homogeneously distributed, to buckle delamination where folded regions focalized the strain, dramatically modifies the arrangement of domain microstructures and this arrangement is not captured by simplistic phase field simulations. Finally, we discuss the mechanisms that drive this heterogeneous distribution of patterns and the possibilities for domain engineering in ferroelastic membranes.

## II. EXPERIMENTAL

Epitaxial films were grown via pulsed laser deposition using conditions reported elsewhere [21]. The X-ray diffraction measurements were performed on a Panalytical X'pert Pro diffractometer (Cu K$\alpha$1), using a parabolic mirror and a (220)Ge 2-bounce monochromator on the incident beam side and a PIXcel position sensitive detector. For the visualization of domain structures, we used four specialized imaging techniques:

### A. Vertical and lateral dual amplitude resonance tracking piezoresponse force microscopy

We used piezoresponse force microscopy (PFM) to study the nanoscale piezoelectric and ferroelectric properties of our samples. In this technique, an oscillating electric field ($V_{ac}$) at frequency $\omega$ is applied to the material by a metallic tip in contact with the sample. The converse piezoelectric effect causes a periodic deformation of the material under the tip, which in turns leads to a periodic bending of the cantilever, also at frequency $\omega$. A lock-in amplifier is used to demodulate the deflection amplitude and phase at $\omega$, where the amplitude signal is proportional to the local polarization magnitude, and the phase signal gives information on the polarization orientation of the material under the tip.

Depending on the material-specific piezoelectric tensor, the deformation of the surface can take place in different directions leading to deflection or torsion of the cantilever. A deflection of the cantilever is a consequence of the out-of-plane deformation (vertical displacement, referred as VPFM), while torsion arises from the in-plane deformation (lateral displacement, referred as LPFM) of the surface of the sample. It is important to note that LPFM, contrary to VPFM, is dependent on cantilever orientation with respect to the polarization axis, thus LPFM measurements must be done at least at two orthogonal orientations of the sample with respect to the cantilever to determine approximately the in-plane polarization direction.

Here we perform the PFM measurements using an Asylum Research MFP-3D atomic force microscope in Dual AC Resonance Tracking (DART) PFM [22] to obtain both vertical and lateral signals. In this technique, the resonance peak is tracked by applying the $V_{ac}$ through the tip-sample system at two frequencies simultaneously, typically 5-15 kHz apart and centered around the contact resonance frequency. Two lock-in amplifiers track the cantilever oscillation amplitude and phase at each of the two drive frequencies and a feedback loop dynamically controls the central frequency by using the difference in amplitudes A2 - A1 as an error signal. The cantilevers used were PPP-EFM (nominal elastic constant of ~2.8N/m).

## B. Channeling-contrast backscattered electron microscopy

Scanning electron microscopy (SEM) has been demonstrated as a powerful non-destructive tool for the observation of domains in ferroelectrics [23]. Backscattered Electron (BSE) imaging, provides channeling contrast sensitive to crystallographic orientation and, therefore, to the misorientation between ferroelastic domains as well as to localized strain. Changes in channeling efficiency translate into electron yield variations (contrast) allowing to precisely map the domain microstructure [24]. In the present work, the domain microstructure of wrinkles was analyzed by SEM using an FEI Magellan 400L XRSEM microscope in the BSE mode, at acceleration voltages comprised between 3 – 5 kV, probe currents in the range 0.1 – 0.4nA, and a working distance of 5 mm.

## C. Polarized micro Raman spectroscopy

Polarized Raman has been proved useful to image the spatial distribution of domains [25] and structural phases in $BaTiO_3$ single crystals [8]. Here we used a micro Raman Spectroscopy setup (RENISHAW Wire Raman microscope), which scans a laser (λ=532nm) over the sample in steps of 150 nm while collecting and measuring the energy of the inelastic scattered photons after their interaction with the specimen's lattice. We used a x100 objective and an exposure time of 1 second per pixel. The measurement configuration was $Z|X\bar{X}|Z$, which means that the incident light came from the Z axis (perpendicular to surface) and polarized along the X-axis (parallel to surface), and the scattered light analyzed is also polarized along the X-axis and collected along the Z -axis (same as incident light). We set the X-axis perpendicular to the direction of the wrinkle.

## D. Scanning transmission electron microscopy

Cross-sectional TEM lamellae were prepared by Ga+ Focused Ion Beam procedures using a Helios 650 de Thermo-Fisher FIB. Atomic-resolution high angle annular dark field scanning transmission electron microscopy (HAADF-STEM) images were acquired using a double-aberration corrected Thermo Fisher Scientific Spectra 300 STEM and a probe-corrected Thermo Fisher Scientific Titan Low Base 60-300, both equipped with CEOS aberration correctors and operated at 300 kV.

## III. RESULTS AND DISCUSSION

### A. Wrinkle formation and morphology

An epitaxial 86 nm (001)-oriented $BaTiO_3$ film was grown on a (110) $GdScO_3$ substrate with a 7 nm buffer water-soluble layer of $Sr_3Al_2O_6$. Upon immersing the film in water, the $Sr_3Al_2O_6$ dissolves [26] and the bond between the $BaTiO_3$ film and the $GdScO_3$ substrate is removed. The film was attached to a polydimethylsiloxane (PDMS) support before etching, and then transferred to a platinum-coated silicon substrate using a dry stamp transfer method [27] [**Fig. 1a**]. During the etching process, when the $BaTiO_3$ film separates from the growth substrate, we observe wrinkle patterns with constant periodicity, induced by the elastic mismatch between PDMS and $BaTiO_3$ (see Suppl. Info. **Fig. S1**). Upon transferring the $BaTiO_3$ films to a rigid silicon substrate by contacting the PDMS stamp with the recipient substrate and slowly peeling off the PDMS at relatively low temperature (70 °C), most of the membrane flattens and some cracks appear; however, we also identify many regions where the membrane remains buckled after the transfer process [**Fig. 1b**]. The morphology of these corrugations on the rigid substrate differs from the original sinusoidal undulations observed on PDMS: on the flat surface, the membrane adopts a shape closer to that of buckle delamination, indicating that during transfer there is a change in the film geometry and in the strain distribution. For simplicity, we will still refer to these buckle delaminations as "wrinkles".

Using AFM, we measured the profile of these membranes across the wrinkles, [inset in **Fig. 1b**]. We observe that the shape profile can be clearly divided in four regions: a transitional concave region between the unwrinkled sheet and the lateral flat slope, where the film starts to separate from the Si substrate, (the "foot"); a straight part (the "slope"); a transitional region between the slope and the convex region at the apex of the wrinkle, (the "neck"); and the apex (the "crest"). This shape can be fitted by the equation representing an Euler buckling deformation [28]:

$$y(x) = \frac{h}{2}\left[1 + \cos\frac{2\pi x}{w}\right] \qquad 1$$

where *h* and *w* are the wrinkle height and width, respectively. We can estimate the strain accumulated at the crest of these wrinkles using the formula [29]:

$$\varepsilon \sim \pi^2 h t/(1 - \sigma^2)w^2 \qquad 2$$

where $t$ is the membrane thickness and $\sigma$ is the Poisson's ratio of BaTiO$_3$ (0.35 [30]). We obtain a strain value of 1.1%, which is remarkably close to the tetragonal strain of BaTiO$_3$ at room temperature [31]. We can also estimate the curvature radius at the crest and foot of the wrinkle, and translate it into a strain gradient, considering the thickness of the membrane. In both regions, we obtain a relatively modest value of $2 \times 10^5$ m$^{-1}$, which is 2 orders of magnitude lower than the strain gradients observed in 5-nm-thick buckled SrTiO$_3$ and BiFeO$_3$ membranes [32] or in 50 nm BaTiO$_3$ membranes forcedly wrinkled via thermal mismatch with a PDMS support [18]. Therefore, flexoelectric effects are expected to be comparatively lower in our membrane.

**B. Ferroelastic domains and hierarchical domain structures at the wrinkles**

X-ray diffraction shows that the crystallinity of the films is maintained after the transfer process [**Fig. 1c**]. The position of the 002 diffraction peak of BaTiO$_3$ in the epitaxial films indicates the out-of-plane lattice expansion of the tetragonal BaTiO$_3$ unit cell due to the compressive strain induced by the substrate. Thickness fringes around the peak are a signature of the high crystalline quality. In the film transferred to silicon, this peak shifts to larger diffracting angle (smaller lattice parameter), coinciding with the position expected for a *c*-oriented BaTiO$_3$ crystal free of strain, and thus indicating a relaxation of the initial epitaxial strain. We also identify a small shoulder at higher angles, suggesting the presence of a small proportion of in-plane (*a*-oriented) domains in the sample. We can in fact detect some evidence of these in-plane domains within the wrinkles by observing the birefringence contrast of the membrane in the optical microscope [**Fig. 1d**]: placing the sample between crossed polarizers, adjusted at 45º from the main crystallographic axes, we detect stripe-shape domains parallel to [100] directions, on both slopes of the wrinkles. Since *c* domains are not expected to give birefringence contrast when observing the film along the [001] axis, these stripes can be ascribed to *a*-domains.

Given that the domain sizes in the wrinkles are within the limit of optical resolution, we performed polarized Raman microscopy maps, using the configuration described in the experimental section and shown in **Fig. 2a**. We mapped an area (~35 × 15 μm) around a wrinkle closely aligned to the <100> orientation of the membrane, shown in the optical image in **Fig. 2b** top panel. Cluster analysis (by k-means clustering) of the acquired spectra in the map allows us to identify three main regions with similar spectral features [**Fig. 2c**]. We focus our analysis on the Raman peaks observed between 450 cm$^{-1}$ and 800 cm$^{-1}$, which are less affected by background intensity subtraction. In this spectral region, both transversal (TO) and longitudinal (LO) optical modes are observed. Cluster 1 corresponds to the flat membrane area and shows a very similar spectral shape to cluster 2, which is observed in the side-slopes of the wrinkle. The main difference between these two spectra is a small shift towards higher frequencies of the peak around 720 cm$^{-1}$ for the spectrum corresponding to cluster 2, likely indicative of strain contrast between these regions, whereas the spectra in cluster 3 are significantly different from the other two. The areas showing the spectral features of cluster 3 are concentrated in the foot and crest of the wrinkle (convex and concavely curved, respectively), and some stripes within the slope connecting the crest and the foot (**Fig. 2b** middle panel). According to previous interpretations of Raman spectra in BaTiO$_3$, the intensity of the Raman modes near 720 cm$^{-1}$ can be used as a proxy for the orientation of ferroelectric domains, since the E(LO) modes are practically muted in the presence of in-plane polarization (*a* domains), whereas intensity is maximized for out-of-plane polarization (*c* domains) [25]. Mapping the integrated intensity of this *c*-polarized mode, we observe a nearly constant intensity in the whole mapped area that goes to zero at the regions that were identified with Cluster 3 (**Fig. 2b** bottom panel). These data suggest that the polarization of the sheet changes from out-of-plane in the flat areas (both in the unwrinkled part and in the side slopes of the wrinkle) to in-plane at the curved regions (crest and foot), with some *a* domains crossing the straight slope and connecting the curved regions.

These results are already surprising for two reasons: (i) one may expect that the flat areas, which are strain-free, would prefer to have their polarization in-plane to reduce depolarizing fields. The fact that polarization is out-of-plane implies that there is either sufficient screening to prevent the flop from the initial epitaxially-induced *c*-axis polarization to *a*-axis polarization, or that the *c*-axis polarization is stabilized by antiparallel 180° domains, and (ii) while in-plane polarization is expected in the wrinkle apex, due to the curvature-induced tensile strain [18], this strain is compressive in the concave region at the feet of the wrinkle, and one would therefore expect out-of-plane polarization in that region, yet the Raman intensity map shows none. Perhaps a more surface-sensitive and/or higher-resolution probe and can determine whether there is a difference in domain configuration between the crest and the feet of the wrinkle. We therefore turn to piezoresponse force microscopy (PFM).

We first used DART-PFM in order to explore the domain patterns at the nanoscale by monitoring the local electromechanical response of the membrane. The vertical PFM signal measured in an area within the same wrinkle explored in Raman [**Fig 3a**] shows faint line contrasts parallel to the wrinkle axis, coinciding with the transitional regions from the unwrinkled sheet to the flat slopes, *i.e.* the foot, and from the slope to the crest, namely the neck. In addition, one observes some traverse lines crossing the slopes perpendicularly to the wrinkle axis, analogous to those visualized by Raman spectroscopy in Fig. 2(b). The lateral PFM signal obtained with the cantilever oriented parallel to the axis of the wrinkles [**Fig 3b**], reveals domains showing a 180° phase contrast at the crest of the wrinkle, indicating the presence of antiparallel $a_1$ domains in this region. However, no clear domains are identified for this scan resolution and for this sample-cantilever orientation at the foot and slope. Once the sample is rotated 90º with respect to the cantilever orientation [**Fig. 3c**], the contrast between the crest and the rest of the wrinkle mostly vanishes, except for the marked contrast with the needle-shaped domains, which extend across the slopes, connecting the crest with the foot, as observed previously in Raman. This suggests that the polarization in these needle domains is rotated by 90º with respect to the domains in the crest ($a_2$ domains).

When performing high resolution scans around the needle domains, we observe a 180º phase contrast contained within the needles, indicating the nucleation of domains within domains. The maximum phase contrast is obtained with the cantilever perpendicular to the wrinkle axis -and parallel to the needle- [**Fig 3d**]. This contrast vanishes when rotating the sample 90º [**Fig 3e**], confirming that the domain microstructure within the needles corresponds to antiparallel $a_2$ domains with in-plane polarization parallel to the wrinkle axis (perpendicular to the needle). Looking more closely at the amplitude images, in these high-resolution images we observe some line patterns in the curved regions (foot and neck) that were not identified previously in the larger scans. Given the orientation of these stripe patterns, parallel to the [010] wrinkle direction, we can associate them to $a_1/c$ domains. In the slopes, however, they rotate 90º forming $a_2/c$ domains. The faint signal in the amplitude images, and the lack of contrast in the phase, are likely due to the small size of these nanodomains, but could also be due to the convoluted contribution between surface and buried layers. Thus, although PFM is mostly sensitive to surface polarization, the underlying material may still have a contribution to the piezoresponse in such thin films which precludes a straightforward interpretation of the data.

With the information gathered so far, we can draw a schematic picture of the multiple domain structures identified in the wrinkles (**Fig. 3f**). Whereas the PFM signal at the crest is governed by the 180º domain configuration of traverse $a$-polarized domains ($a_1$ in the schematic), in the bounding "neck" and in the "foot" transitional regions we observe a superimposed signal coming from $a_1/c$ domains extending parallel to the wrinkle axis, and traverse $a_2/c$ nanodomains confined within the $a_1/c$ bundles. Those two domain types, $a_1$ and $a_2$, will be hereafter referred to as primary and secondary structures, respectively. The formation mechanisms of such hierarchical structures are discussed in next section.

**C. Effect of wrinkle curving on the ferroelastic domain pattern**

From a mechanical point of view, ferroelastic sheets can be considered equipped with a system of "folding" directions determined by their point symmetry. An analogy between folds and twins here arises from the fact that a permissible domain wall localizes the twin shear strain on a given plane orientation at a minimum energy cost. This is what occurs in a wrinkle-to-fold transition above a critical strain [33]. In BaTiO$_3$ (point group symmetry 4mm), permissible domain walls lie on {110} planes [34]. To rationalize the observed wrinkle orientations and the distribution of domains along the wrinkle profile we note that wrinkling causes traverse shear stress. This shear acts on the surface of the sheet and can, therefore, couple with the twin shear of *a/c* domains, this way favoring the formation and self-organization of {011} domain walls along the in-plane <100> directions. Note that the twin shear of $a_1/a_2$ domains is incompatible with the wrinkle symmetry and therefore those domains cannot be formed.

The observed distribution of …$a_1/c/a_1/c$… bundles concentrating the curvature along the crest, neck and foot regions indicates that the formation of wrinkles in free sheets encompasses a specific, highly reproducible, primary architecture, not captured by atomistic simulations which constrain the wrinkle profile to a sine-wave function [18]. Our results show that self-organization of *a/c* domain walls in free sheets adds a morphological degree of freedom to be considered in a wrinkling based approach to domain engineering. This degree of freedom controls domain density and localization.

Considering the absolute value of the strain gradient at the crest, neck and foot regions, ~2 × 10$^5$ m$^{-1}$ (see section III.A, and **Fig. S2**), for a ~90 nm thickness, the strain difference between the top and bottom surfaces in those regions, ~2%, doubles the tetragonal distortion of BaTiO$_3$ [31]. This gradient cannot be accommodated just by *a/c* domain walls and, therefore, the observed domain microstructures should be accompanied by elastic deformation giving rise to dipole strength variations and rotations, as observed in wrinkled films bonded to viscoelastic substrates [18]. To directly access the atomic structure of the wrinkle, an electron transparent cross sectional sample was prepared by focused ion beam (**Fig. S2a**). As shown in **Figs. S2b-d**, neither lattice images nor strain analysis revealed the occurrence of any domain microstructure but, instead, the bending strain was observed to be elastically accommodated by linear strain gradients in the crest, neck and foot regions. For the crest, this result is consistent with the occurrence of ±$a_1$-domains siting on the convex top of the wrinkle, in agreement with the PFM and Raman analysis. However, for the neck and foot, the atomic structure of the electron transparent sample (actually, a bend wire), lacks the fine domain microstructure identified by PFM and SEM. We interpret this behavior in terms of the lability of such structures: the vanishing order parameter along the strain gradient causes ferroelastic domains to be confined in the neighborhood of convex surfaces, making them labile against changes in boundary conditions. During the preparation of a thin specimen, the aspect ratio (axial length of the wrinkle divided by the sheet thickness), is dramatically reduced from infinite to ~1. We believe that the associated increase in surface-to-volume ratio in the resulting wire-like sample increases the surface contribution to strain relaxation causing the suppression of adaptive microstructures.

To better explore the hierarchical distribution of primary and secondary domains, we surveyed the sample using the channeling contrast from the BSE signal in SEM. We identify a variable density of ferroelastic domains in different wrinkles or even in different parts of the same one, correlated with the deviation of its orientation with respect to the main crystallographic axis. **Fig. 4a** illustrates the overall effect of in-plane curvature and misorientation on the domain microstructure. It can be readily observed that while the region aligned with the [010] direction is only accommodated by primary domains, as the deviation from the preferred orientation increases, the microstructure becomes plagued by additional features: (*i*) domain wall steps, (*ii*) needles associated to the steps, parallel to the vertical [010] direction (marked by a green box), (*iii*) traverse secondary domain structures grouped in bundles of varying length and confined inside the neck and foot regions (seen inside yellow rectangles), and (*iv*) longer traverse domains spanning through the slopes, as the one arrowed in the upper left of the wrinkle.

**Fig. 4b** is a magnified view of the upper region of the same wrinkle (for reference, the yellow arrow indicates the same needle domain as marked in (**a**). Careful inspection of the wrinkle crest reveals very weak traverse contrasts which we attribute to ±$a_1$ 180º domains identified by PFM (**Fig. 3**). At each side of the crest within the neck regions, the high-resolution BSE image allows to discern a rich microstructure combining features (*i*) through (*iv*) mentioned in the preceding paragraph. Wrinkle misorientation forces … $a_1/c/a_1/c$ … primary bundles away from their permissible {110} orientation (*i.e.*, away from the in-plane <100> directions). Consequently, they become stepped keeping long ledges parallel to the preferential in-plane <100> orientation, as indicated by dashed lines in **Fig. 4b**. The steps themselves, on the other hand, should sustain the misfit strain resulting from the difference between the tetragonal *a* and *c* lattice parameters, causing local strain concentration. Their contrast, instead, is observed in most cases to fade along needles of varying length, as shown within the green box in **Fig. 4a** and vertical dashed lines in (**b**). The left panel in **Fig. 4c** schematizes $a_1$ and *c* needle formation at misfitting $a_1/c$ and $c/a_1$ steps, respectively. As needles minimize the step interface area and hence strain energy, their fading suggests a smooth decrease of the needle cross section towards their tip.

Under suitable imaging conditions, *i.e.* where contrast is favored by the local crystal orientation, it is possible to discern the overlapped step/needle patterns through the sampled volume (see BSE image in **Fig. 4c**, right panel). The schematics in the central panel is a guide for the interpretation of the BSE image contrast. The image shows that traverse $a_2$ nanodomains preserve the contrast of the primary structure suggesting they are polygonised into a mosaic pattern following the orientation of the intertwined …$a_1/c/a_1/c$… bundle. Therefore, the intersection of primary and secondary structures generates, in turn, new subdomains building a complex topological pattern.

Besides the generation of dense stepped profiles and needles along … $a_1/c/a_1/c$ … bundles, in-plane curvature plagues the microstructure with traverse nanodomains ($a_2$) confined within the softer neck and foot regions (see yellow rectangles in **Fig. 4a**). Such nanodomains form bundles of varying length with a remarkably homogeneous apparent domain width of ≈200 nm and are thought to accommodate the bending strain in a similar way as primary bundles. Their uneven distribution along with associated strong strain contrasts suggests domain pinning, probably due to oxygen vacancies [35, 36], which would indicate domain motion at some stages of the curving process.

**Fig. 4b** clearly shows that longer needle domains exclusively grow from such secondary bundles, keeping their 90º rotated polarization, as demonstrated by our PFM analyses. These larger needles may even span through the whole slope of the wrinkle forming larger domains connecting the foot and neck regions and constitute the only adaptive degree of freedom within the slopes of the wrinkle. As can be observed in **Fig. 4b** (see inset) BSE images reveal an inner domain structure within large traverse needles. According to our PFM analysis (**Fig. 3**), such contrast would correspond to 180º secondary domains, as labelled ($\pm a_2$) in the Figure.

## IV. CONCLUSION

Our results show that lateral curving causes a self-organized multiscale domain pattern in wrinkled ferroelectric/ferroelastic BaTiO$_3$ free sheets (*i.e.*, films free of any compliant or rigid substrate). To rationalize the pattern formation mechanism, one must first consider the wrinkling behavior of a straight wrinkle in a free-sheet: in this case, $a_1/c$ primary domain walls self-organize to accumulate curvature along two narrow regions, one adjacent to the wrinkle crest (neck), and one along the transition between the slope and the unwrinkled sheet (foot). This way, neck and foot regions provide softer hinge-like regions rendering the wrinkle a folding character, leaving flat lateral slopes in contrast with wrinkles *sensu stricto* (*i.e.*, adhered to a compliant substrate), which tend to develop a sinusoidal profile as captured by atomistic simulations [18].

Curving such a domain architecture combines two types of response: (*i*) Shear stresses drive a hierarchical distribution of traverse nano-domains with 90º rotated in-plane polarization ($a_2$), forming bundles confined along the neck and foot regions. Such nano-domains eventually span through the slopes of the wrinkle to relax residual strains in the slopes, and (*ii*) the deviation of $a_1/c$ domain walls from their permissible (equilibrium) orientation causes their decomposition into steps and longer ledges keeping the original crystallographic planes. The resulting hierarchical domain microstructure can, therefore, be understood by the superposition of two curvatures, a primary one about the axis of the wrinkle, and a secondary one about an axis perpendicular to the sheet. Primary and secondary strain fields each drive the self-organization of their corresponding domain orientations, $a_1$ and $a_2$, with the particularity that the latter become hierarchically confined within the formers.

These findings set the grounds for a deterministic control of multiscale dense domain and domain-wall patterns in ferroelectric/ferroelastic free sheets by taking advantage of the rich coupling between domain formation and bending. One could think, for example, on nanofabrication strategies for synthesizing wrinkle networks on ferroelectric oxide membranes [37], allowing to fix particular orientations of these wrinkles that promote dense domain structures as hinted in our study, or use flexible and stretchable substrates to mechanically control the wrinkle density, orientation and amplitude [38], therefore actively tuning domain structures and properties through wrinkle manipulation, which one may call "wrinkletronics".

In summary, we have studied the morphology and domain structure of wrinkles in a thin ferroelectric $BaTiO_3$ free sheet transferred to silicon. We have found that these wrinkles concentrate strains in the order of 1%, which are accommodated via ferroelastic domain structures to adapt to curvature and local strain, by developing a complex domain structure in which a hierarchy of ferroelastic domains, mainly $a/c$, accommodates both the curvature across the thickness of the wrinkle and the shear strain imposed by the planar deviations from the main crystallographic axes of $BaTiO_3$, both yielding regions of coexistence between multiple microstructures and high density of domains. Our study thus reveals the relevance of competing strains and ferroelastic domain formation in the microstructure of bent freestanding oxide membranes, which needs to be considered when studying the properties in these quasi-2D materials and when designing functionalities exploiting the flexibility of oxide membranes.

**ACKNOWLEDGMENTS**

D.P. acknowledges funding from 'la Caixa' Foundation fellowship (ID 100010434). The authors acknowledge the Spanish Ministry of Industry, Economy and Competitiveness (MINECO) through Grants no. PID2019-108573GB-C21, PID2022-140589NB-I00, PID2020-112914RB-I00 and PID2023-147211OB-C22 funded by Grant No. MCIN/AEI/10.13039/501100011033. The Micro Raman imaging research was conducted as part of a user project at the Center for Nanophase Materials Sciences (CNMS), which is a US Department of Energy, Office of Science User Facility at Oak Ridge National Laboratory. Authors acknowledge the use of instrumentation financed through Grant IU16-014206 (METCAM-FIB) to ICN2 funded by the European Union through the European Regional Development Fund (ERDF), with the support of the Ministry of Research and Universities, and Generalitat de Catalunya. The ICN2 is funded by the CERCA programme/Generalitat de Catalunya and by the Severo Ochoa Centres of Excellence Programme, funded by the Spanish Research Agency (AEI, CEX2021-001214-S). The ICMAB is funded the Severo Ochoa Centres of Excellence Programme, funded by the Spanish Research Agency (AEI, CEX2023-001263-S). The INMA is funded by the Severo Ochoa Centres of Excellence Programme (AEI, CEX2023-001286-S). The authors acknowledge the European's Union Horizon 2020 research and innovation programme under Grant No. 823713-ESTEEM3. This work was also funded from Regional Gobierno de Aragon through Project Nos. E13_23R, including FEDER funding. The authors also acknowledge the use of instrumentation as well as the technical advice provided by the National Facility ELECMI ICTS, node «Laboratorio de Microscopias Avanzadas (LMA)» at «Universidad de Zaragoza». The authors are grateful to Dr. Kapil Gupta for assistance with the Spectra electron microscope at the Joint Electron Microscopy Centre at ALBA synchrotron (JEMCA).

[a] D.Pesquera, K.Cordero and F.Sandiumenge contributed equally to this work.

**FIGURES**

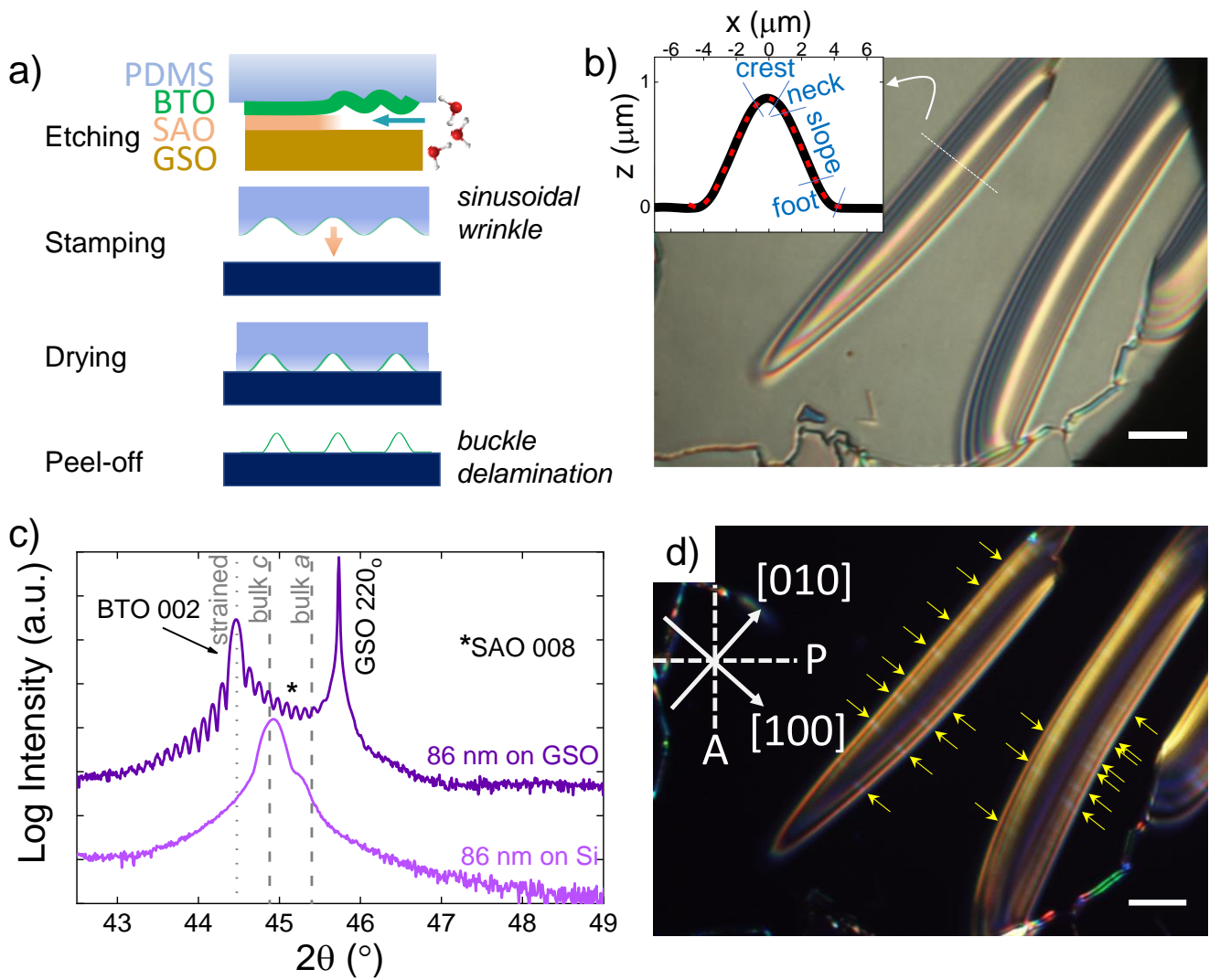

FIG. 1. a) Schematic of film release from growth substrate, formation of sinusoidal wrinkles during etching and transition to buckle-delamination-like wrinkles upon transfer to a rigid substrate. b) Optical image of wrinkles in the BaTiO$_3$ membrane on silicon; inset shows a profile across one of the wrinkles, measured with AFM, dashed red line is a fit to eq.(1). c) $\theta$-$2\theta$ XRD scans the BaTiO$_3$ film as grown on GdScO$_3$ and after release and transfer to Pt-coated silicon d) Same region as b), imaged using crossed polarizers, both at 45° from the main in-plane ⟨100⟩ axes. Arrows signal some of the locations in the wrinkle where contrast due to ferroelastic domains are identified. Scalebars in b and d are 10 μm.

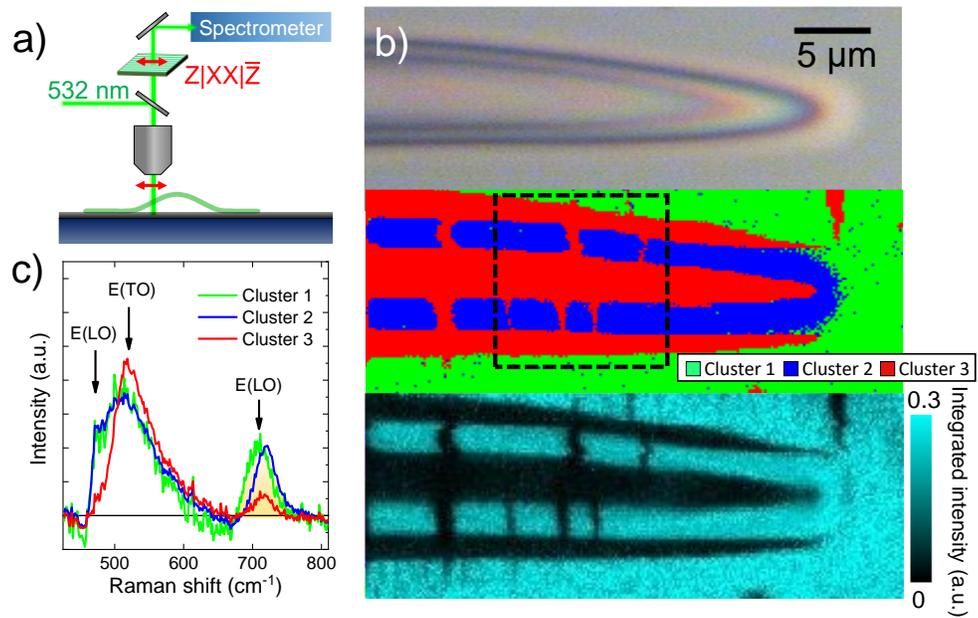

FIG. 2. a) Schematic of the polarized Raman measurement. b) Optical image of the measured wrinkle (top), mapping of clusters with similar spectral features (middle) and mapping of integrated area under E(LO) Raman peak around 720 cm$^{-1}$ (bottom). Dashed square in middle indicates area explored in fig.3. c) Average spectra of the areas corresponding to the three identified clusters around a wrinkle in the BaTiO$_3$ nm membrane on Pt/Si.

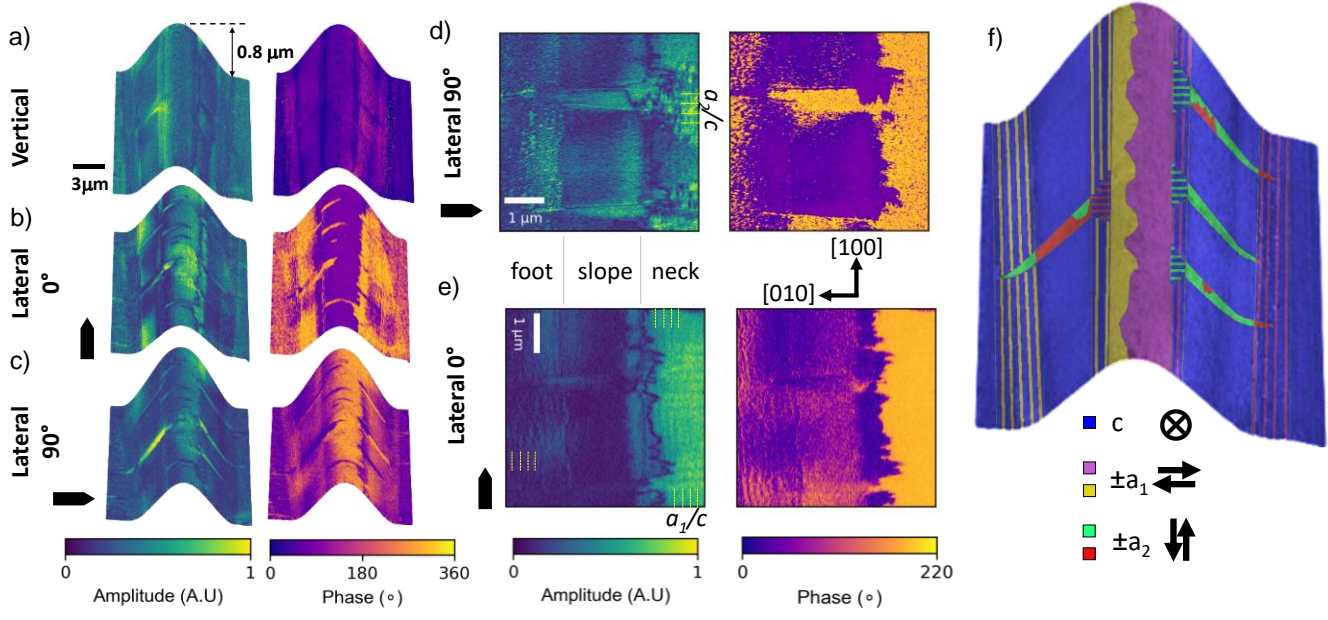

FIG. 3. Vertical (a) and lateral (b-e) PFM amplitude and phase scans taken around a wrinkle in the BaTiO$_3$ membrane. (d,e) are zoomed areas within the wrinkle, around needle domains. Lateral images were taken with cantilever parallel (b,e) and perpendicular (c,d) to the wrinkle axis. f) Schematic of domain pattern distribution as "planar view".

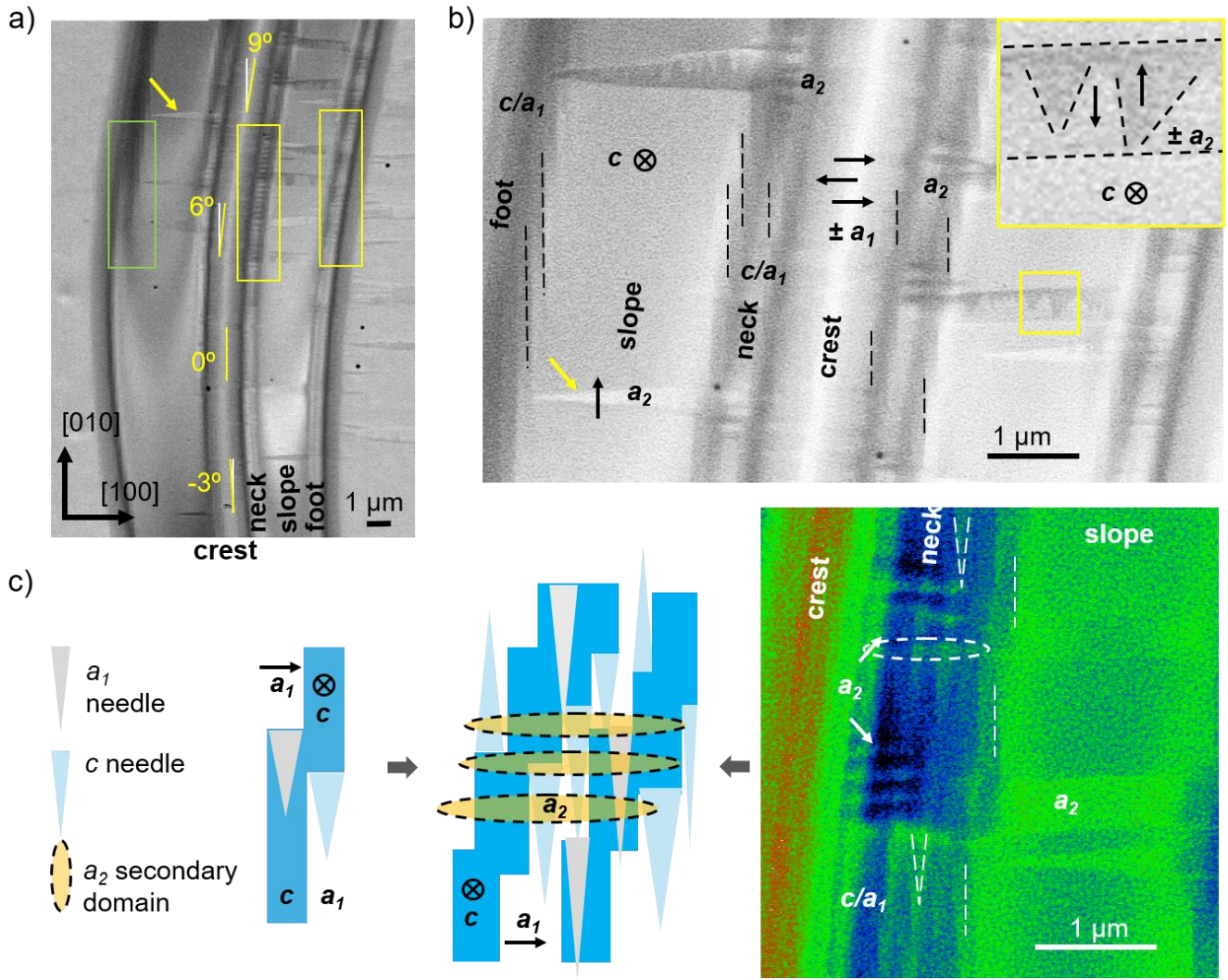

FIG. 4. a) BSE image of a curved wrinkle with varying degrees of deviation from the equilibrium [100] direction. The image shows increased density of secondary (traverse) domain microstructures with increasing deviation angle. The different regions of the prototypical wrinkle are indicated at the bottom. The green window shows needles growing from domain wall steps, and yellow windows show bundles of secondary nanodomains confined within the neck and foot regions. b) Enlarged view of the upper region of **a**) (the yellow arrow points the same needle as in **a**)). Domain notations and polarization vectors (black arrows) are deduced from PFM and Raman microscopies (**Figs. 2,3**); the sign of vectors is arbitrary. Vertical dashed lines are a guide to the eye to show the orientation of $a_1/c$ domain-wall step ledges along the [010] direction. The inset is a magnified view of the yellow boxed area illustrating a domains-within-domain structure formed within traverse needles spanning through the wrinkle slopes. c) *Left panel*: schematics of needle growth from misfitting steps *Right panel*: BSE image of the transition from the neck to the slope region of a misoriented wrinkle, showing in detail the superposition effect of steps and needles along stepped $a_1/c$ domain-walls. False color is used to highlight fine contrast within the neck region. Straight dashed lines are guides to the eye to identify step ledges (vertical lines) and needles. Secondary $a_2$ nanodomains crossing the primary …$a_1/c/a_1/c$… domain bundle are indicated. *Central panel*: Schematical interpretation of the BSE image shown in the right panel.

**REFERENCES**

[1] A. R. Damodaran *et al.*, *Adv. Mater.* **29**, 1702069 (2017).

[2] S. Pandya *et al.*, *Adv. Mater.* **31**, 1803312 (2019).

[3] G. Velarde *et al.*, *APL Mater.* **9** (2021), doi:10.1063/5.0035735.

[4] H. Fu *et al.*, *Nature*. **403**, 281–283 (2000).

[5] S. Park *et al.*, *J. Appl. Phys.* **82**, 1804–1811 (1997).

[6] S. Pandya *et al.*, *Nat. Mater.* **17**, 432–438 (2018).

[7] A. S. Everhardt *et al.*, *Appl. Phys. Rev.* **7**, 011402 (2020).

[8] T. T. A. Lummen *et al.*, *Nat. Commun.* **5**, 3172 (2014).

[9] S. Wada *et al.*, *J. Appl. Phys.* **98**, 014109 (2005).

[10] D. Pesquera *et al.*, *J. Phys. Condens. Matter*. **34**, 383001 (2022).

[11] F. M. Chiabrera *et al.*, *Ann. Phys.* **534**, 2200084 (2022).

[12] S. R. Bakaul *et al.*, *Adv. Mater.* **29**, 1605699 (2017).

[13] Z.-D. Luo *et al.*, *ACS Appl. Mater. Interfaces*. **11**, 23313–23319 (2019).

[14] D. Pesquera *et al.*, *Adv. Mater.* **32**, 2003780 (2020).

[15] P. Gentile *et al.*, *Nat. Electron.* **5**, 551–563 (2022).

[16] G. Dong *et al.*, *Science*. **366**, 475–479 (2019).

[17] C. Guo *et al.*, *Appl. Phys. Lett.* **116**, 152903 (2020).

[18] G. Dong *et al.*, *Adv. Mater.* **32**, 2004477 (2020).

[19] J. Wang *et al.*, *Adv. Sci.* **11** (2024), doi:10.1002/advs.202401657.

[20] H. Mei *et al.*, *Appl. Phys. Lett.* **90**, 151902 (2007).

[21] S. Ganguly *et al.*, *Adv. Mater.* **36** (2024), doi:10.1002/adma.202310198.

[22] B. J. Rodriguez *et al.*, *Nanotechnology*. **18**, 475504 (2007).

[23] R. Le Bihan, *Ferroelectrics*. **97**, 19–46 (1989).

[24] J. F. Ihlefeld *et al.*, *J. Mater. Sci.* **52**, 1071–1081 (2017).

[25] F. Rubio-Marcos *et al.*, *Nat. Photonics*. **12**, 29–32 (2018).

[26] D. Lu *et al.*, *Nat. Mater.* **15**, 1255–1260 (2016).

[27] A. Castellanos-Gomez *et al.*, *2D Mater.* **1**, 011002 (2014).

[28] D. Vella *et al.*, *Proc. Natl. Acad. Sci. U. S. A.* **106**, 10901–10906 (2009).

[29] A. Castellanos-Gomez *et al.*, *Nano Lett.* **13**, 5361–5366 (2013).

[30] A. C. Dent *et al.*, *J. Eur. Ceram. Soc.* **27**, 3739–3743 (2007).

[31] H. F. Kay *et al.*, *London, Edinburgh, Dublin Philos. Mag. J. Sci.* **40**, 1019–1040 (1949).

[32] S. Cai *et al.*, *Nat. Commun.* **13**, 5116 (2022).

[33] D. P. Holmes *et al.*, *Phys. Rev. Lett.* **105**, 038303 (2010).

[34] J. Sapriel, *Phys. Rev. B*. **12**, 5128–5140 (1975).

[35] X. Ren, *Nat. Mater.* **3**, 91–94 (2004).

[36] D. C. Lupascu *et al.*, *J. Am. Ceram. Soc.* **89**, 224–229 (2006).

[37] A. Malachias *et al.*, *ACS Nano*. **2**, 1715–1721 (2008).

[38] J. A. Rogers *et al.*, *Science*. **327**, 1603–1607 (2010).